\documentclass[12pt]{iopart}
\usepackage{graphicx}

\newcommand{\sqrts}{\mbox{$\sqrt{s}$}}

\newcommand{\dAu}{\textit{d}+Au}

\newcommand{\ep}{\textit{e}+\textit{p}}
\newcommand{\eA}{\mbox{\textit{e}+A}}

\newcommand{\pp}{\mbox{\textit{p}+\textit{p}}}

\newcommand{\pA}{\mbox{\textit{p}+A}}
\renewcommand{\AA}{\mbox{A+A}}
\newcommand{\pT}{\mbox{$p_T$}}

\newcommand{\gev}{\mbox{$\mathrm{GeV}$}}

\newcommand{\lumi}{\mbox{$\mathrm{cm}^{-2}\mathrm{sec}^{-1}$}}

\newcommand{\fmc}{\mbox{$\mathrm{fm}/c$}}

\newcommand{\Qss}{\mbox{$Q_s^2$}}

\begin{document}


\title[The Electron Ion Collider]{The Emerging QCD Frontier: The Electron Ion Collider}

\author{Thomas Ullrich}
\address{Brookhaven National Laboratory, Upton, New York, 11973}

\begin{abstract}
    The self-interactions of gluons determine all the unique features
    of QCD and lead to a dominant abundance of gluons inside matter 
    already at moderate $x$. Despite their dominant
    role, the properties of gluons remain largely unexplored.
    Tantalizing hints of saturated gluon densities have been found in
    $e$+p collisions at HERA, and in d+Au and Au+Au collisions at
    RHIC.  Saturation physics will have a profound influence on
    heavy-ion collisions at the LHC. But unveiling the collective
    behavior of dense assemblies of gluons under conditions where
    their self-interactions dominate will require an Electron-Ion
    Collider (EIC): a new facility with capabilities well beyond those
    of any existing accelerator.
    In this paper I outline the compelling physics case for $e$+A
    collisions at an EIC and discuss briefly the status of machine
    design concepts.
\end{abstract}



\section{Introduction}

Quantum Chromodynamics is a cornerstone of the standard model of
physics.  Many phenomena of QCD that are not directly evident from the
Lagrangian have emerged as our knowledge improved over time.  These
include chiral symmetry breaking and confinement, both of which are
now recognized as defining features of the strong interactions.

Lattice gauge and effective field theories have taught us that the
complex structure of the QCD vacuum arises primarily from the dynamics
of gluons with small contributions from the quark sea.  In fact, the
self-interactions of gluons determine all the unique and essential
features of QCD and lead to a dominant abundance of gluons inside
matter.  The lion's share, 98\%, of the mass of nuclear matter is due
to the gluon interactions, that generate the momentum-dependent
constituent quark masses and the nucleons themselves. Despite this
dominance, the properties of gluons in matter remain largely
unexplored.  Since the gluon degrees of freedom are missing in the
hadronic spectrum it requires high-energy probes reaching deep into
hadronic matter in order to explore "glue" experimentally.

The world's premier accelerator facilities focused on QCD, RHIC
and HERA, have increasingly addressed selected aspects of gluon
behavior accessible to them.  Counter-intuitively, high-energy lepton
beams provide the best gluon microscope, by interacting primarily with
electrically charged quarks, in a process known as deep inelastic
scattering (DIS). The gluonic part of the hadronic wave function
modifies the precisely understood electromagnetic interaction in ways
that allow us to infer the gluon properties (\textit{e.g.} through
scaling violation of structure functions). For this inference to be
precise, DIS must be studied over a broad range of energies and
scattering angles.
 
 \begin{figure}[tb]
     \begin{center}
     \includegraphics[width=0.44\textwidth]{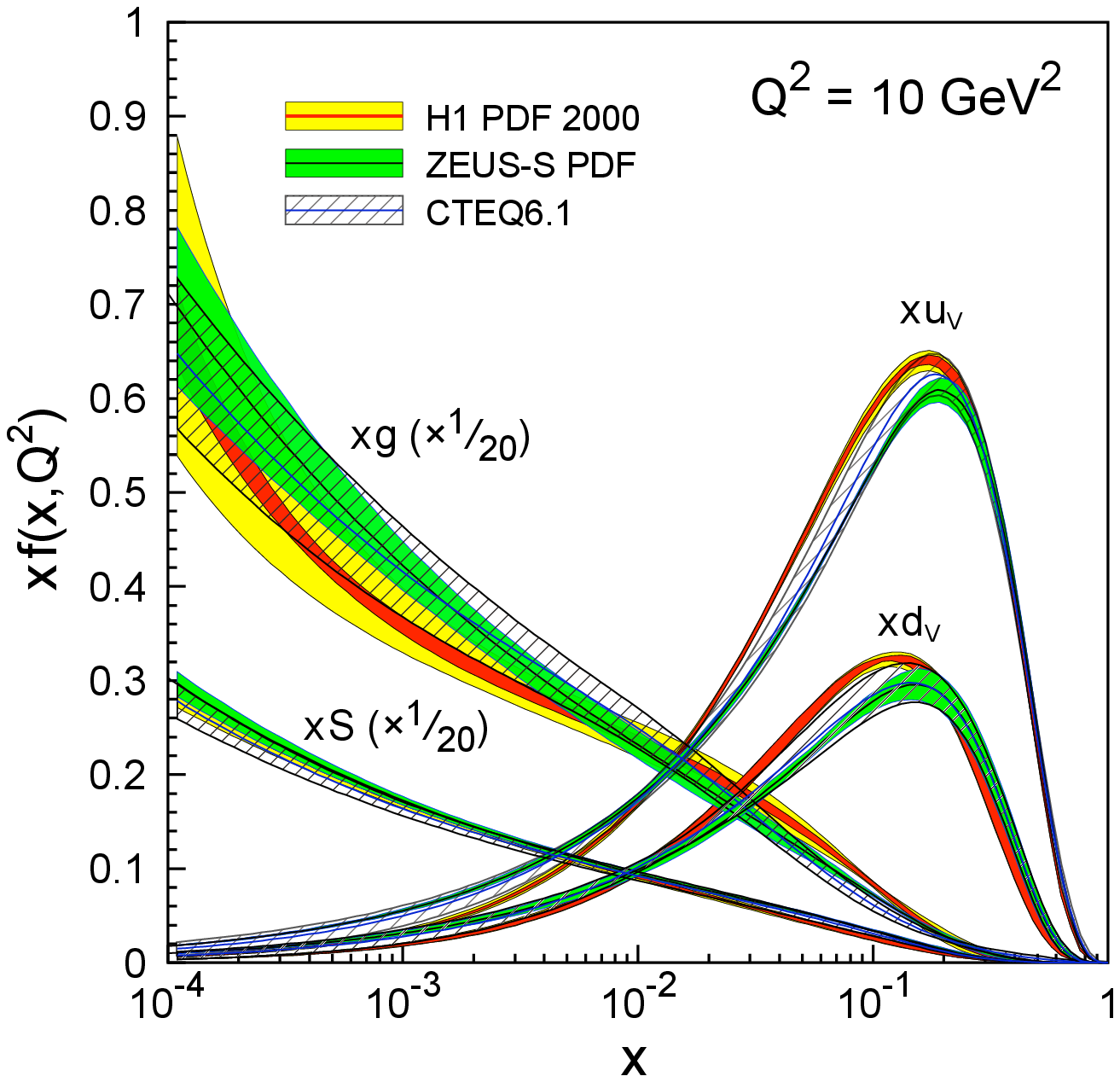}
     \hspace{0.03\textwidth}
     \includegraphics[width=0.46\textwidth]{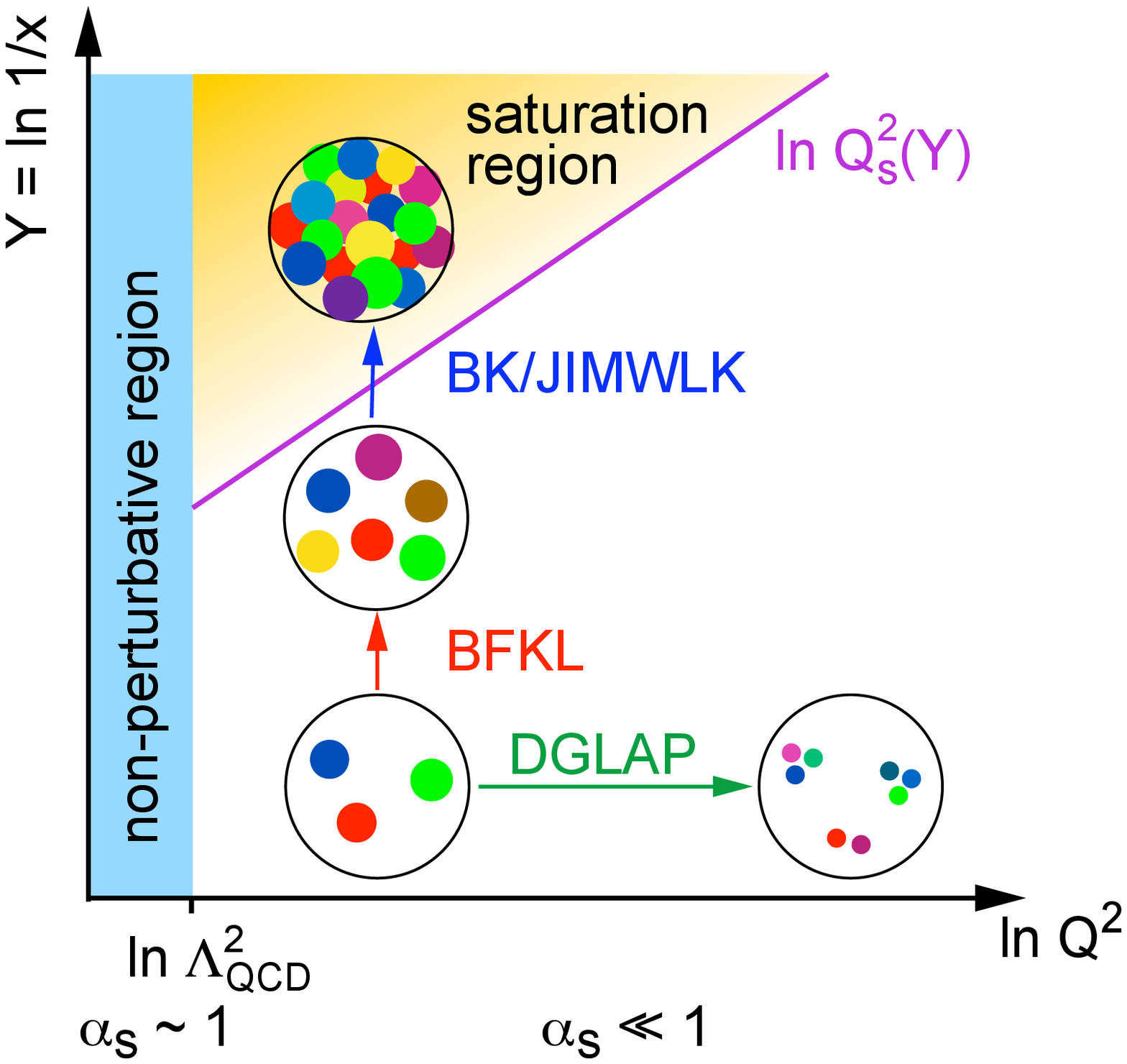}
     \vspace{-5mm}
     \end{center}
     \caption{\textbf{Left:} Gluon, valence, and sea quark momentum
         distributions in the nucleon obtained from a NLO DGLAP fit to
         the proton structure function $F_2$ measured at HERA (from
         \cite{Saxon:2007zz}). Note that the gluon and sea quark
         distributions are scaled down by a factor of 20. Their rapid
         rise at low-$x$ as derived in a linear
         evolution scheme will ultimately lead to a violation of the unitary
         bound. \textbf{Right:} Regions of the nuclear wave function
         in the $\ln 1/x$ versus $\ln Q^2$ plane. The line indicating
         the saturation regime reflects a line of constant gluon
         density. It represents not a sharp transition but indicates the approximate 
         onset of saturation phenomena. }
    \label{fig:pdf}
\end{figure}

Only in DIS of electrons from nuclei, can one
measure directly the momentum fraction $x$ carried by the struck
parton before the scattering and the momentum $Q$ transferred to the
parton in the scattering process.  The invariant cross-section in DIS
can be written as:
\begin{eqnarray*}
\frac{d^2 \sigma^{eA \rightarrow eX}}{dx dQ^2} =
\frac{4 \pi \alpha^2_{e.m.}}{xQ^4} \left[ \left(1-y+\frac{y^2}{2} \right )
    F^A_2(x,Q^2) - \frac{y^2}{2} F^A_L(x,Q^2) \right]
\end{eqnarray*}
where $y$ is the fraction of the lepton's energy lost in the nuclei
rest frame.  The fully inclusive structure functions $F_2^A$ and
$F_L^A$ offer the most precise determination of quark and gluon
distributions in nuclei.  The former is sensitive to the sum of quark
and anti-quark momentum distributions; at small $x$, these are the sea
quarks. The latter is sensitive to the gluon momentum distribution. 
While $F_2$ was extensively studied for protons at HERA \cite{Saxon:2007zz} our knowledge
on $F_L$ is rather limited since it requires measurements at varying $\sqrt{s}$.
The ongoing analysis of the last HERA run in 2007, performed at a lower energy, will
provide a first glimpse of $F_L$ of the proton.

DIS experiments of electrons off protons at the HERA collider have
shown that, for $Q^2 \gg \Lambda_{\rm QCD}^2$, the gluon density grows rapidly with
decreasing $x$ (see Fig.~\ref{fig:pdf} left).  For $x < 0.01$ the
proton wave function is predominantly gluonic. DIS experiments with
nuclei have established that quark and gluon distributions in nuclei
exhibit shadowing; they are modified significantly relative to their
distributions in the nucleon. However, in sharp contrast
to the proton, the gluonic structure of nuclei is not known for
$x<0.01$. At large $x$ and at large $Q^2$, the properties of quarks
and gluons are described by the linear evolution equations DGLAP
\cite{Gribov:1972ri,Altarelli:1977zs,Dokshitzer:1977sg} (along $Q^2$)
and BFKL \cite{Kuraev:1977fs,Balitsky:1978ic} (along $x$). The rapid
growth in gluon densities with decreasing $x$ is understood to follow
from a self similar Bremsstrahlung cascade where harder, large $x$,
parent gluons successively shed softer daughter gluons.  Gluon
saturation is a simple mechanism for nature to tame this growth. When
the density of gluons becomes large, softer gluons can recombine into
harder gluons.  The competition between linear QCD Bremsstrahlung and
non-linear gluon recombination causes the gluon distributions to
saturate at small $x$. The non-linear, small-$x$ renormalization group
equations, JIMWLK
\cite{JalilianMarian:1997gr,JalilianMarian:1997dw,Iancu:2000hn,Ferreiro:2001qy}
and its mean field realization BK
\cite{Balitsky:1995ub,Kovchegov:1999ua}, propagate these non-linear
effects to higher energies leading to saturation.  The onset of
saturation and the properties of the saturated phase are characterized
by a dynamical scale $Q_s^2$ which grows with increasing energy
(smaller $x$) and increasing nuclear size A.
 
The nucleus is an efficient {\bf amplifier} of the universal physics
of high gluon densities. Simple considerations suggest that $Q_s^2
\propto (A/x)^{1/3}$. This dependence is supported by detailed studies
\cite{Kowalski:2007rw,Kowalski:2003hm}.  Therefore, DIS with large
nuclei probes the same universal physics as seen in DIS with protons
at $x$'s at least two orders of magnitude lower (or equivalently an
order of magnitude larger $\sqrt{s}$). When $Q^2 \gg Q_s^2$, one is in
the well understood ``linear" regime of QCD. For large nuclei, there is
a significant window at small $x$ where $Q_s^2 \gg Q^2 \gg
\Lambda_{\rm QCD}^2$ and where one is in the domain of strong
non-linear gluon fields (see Fig.~\ref{fig:pdf} right).

The intensity of the chromo-electric and chromo-magnetic fields in the
strong gluon field regime is of order $\cal{O}$($1/\alpha_S$), where
the asymptotic freedom of QCD dictates that the fine structure
constant $\alpha_S(Q_s^2) \ll 1$. These fields are therefore the
strongest fields in nature! Remarkably, the weak coupling suggests
that the onset and properties of this regime may be computed
systematically in a QCD framework.  The high occupation numbers of
gluons ensures that their dynamics are classical and their piling up
at a characteristic momentum scale ($Q_s^A$) is reminiscent of a
Bose-Einstein condensate.  Dynamical and kinematic considerations have
led to a suggestion that the matter in nuclear wave functions at high
energies is universal and can be described as a Color Glass Condensate
(CGC)~\cite{Iancu:2003xm,Weigert:2005us}.
 
While hints for saturation phenomea have been obtained at HERA and
RHIC, getting to the heart of the matter, will require a new facility
with capabilities well beyond those of any existing accelerator, an
Electron Ion Collider (EIC) \cite{eicc}.

Such a facility will provide definitive answers to compelling physics
questions essential for understanding the fundamental structure of
hadronic matter, and allow precise and detailed studies of the nucleus
in the regime where its structure is overwhelmingly determined by the
gluons.


\section{Connections to RHIC and LHC} 

Measurements over the last six years in heavy-ion collision
experiments at RHIC indicate the formation of a strongly coupled
plasma of quarks and gluons (sQGP)
\cite{Adcox:2004mh,Adams:2005dq,Back:2004je,Arsene:2004fa,Muller:2007rs,Shuryak:2007zz,Gyulassy:2004zy}.
While the evidence for this picture is compelling, there is still no
quantitative framework to understand all the stages in the expansion
of the hot and dense matter. The EIC can contribute to a better
understanding of the dynamics of heavy-ion collision-from the initial
formation of bulk partonic matter to jet quenching and
hadronization that probe the properties of the sQGP.

\textbf{Initial Conditions for the sQGP}.  Understanding the
mechanisms that lead to {\it rapid equilibration} in heavy-ion
collisions is perhaps the major outstanding issue of the RHIC program.
Hydrodynamic modeling of RHIC data suggests that the system achieves
nearly complete thermalization no later than 1\,\fmc\ after the
initial impact of the two nuclei.  These models are very sensitive to
the initial pre-equilibrium properties of the matter (often referred
to as ``Glasma" \cite{Raju_QM2008}). At present there is no first
principle understanding of thermalization in QCD.  A simple picture of
the initial state formation suggests that thermalization is driven by
low-$x$ gluons with $k_T^2 < Q_s^2$ freed on first impact. In this
case, the saturation scale ($Q_s$) defines the scale for the formation
and thermalization of strong gluon fields from the nuclear wave
functions~\cite{Krasnitz:2000gz,Kharzeev:2000ph,Kharzeev:2001gp,Baier:2000sb}.
Any substantial progress in the understanding of the thermalization
process that gives rise to the sQGP will require the profound
knowledge of the momentum and spatial distribution of gluons in nuclei
$G_A(x, Q^2, b)$.

\begin{figure}[tb]
     \begin{center}
     \includegraphics[width=0.45\textwidth]{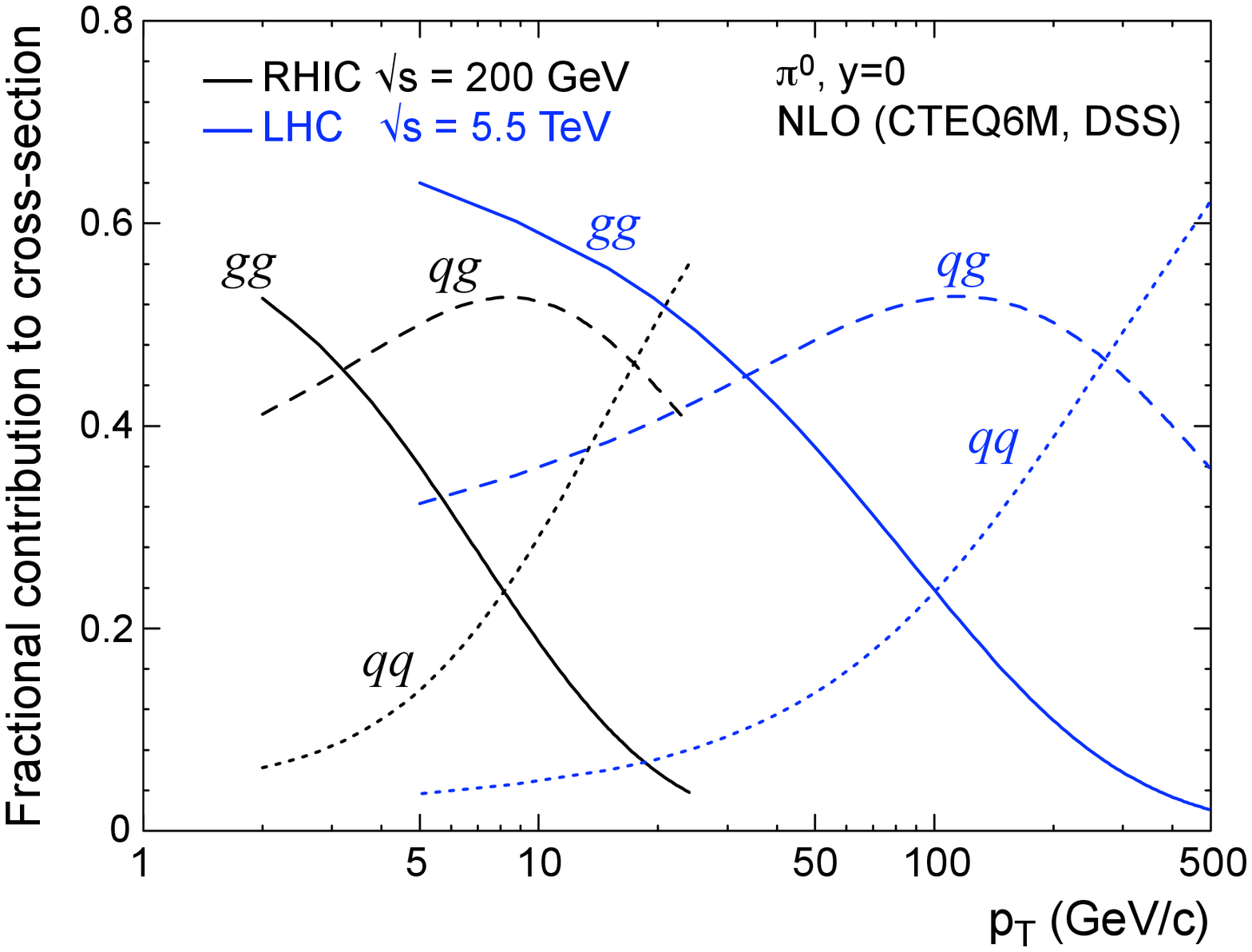}
     \hspace{0.05\textwidth}
     \includegraphics[width=0.47\textwidth]{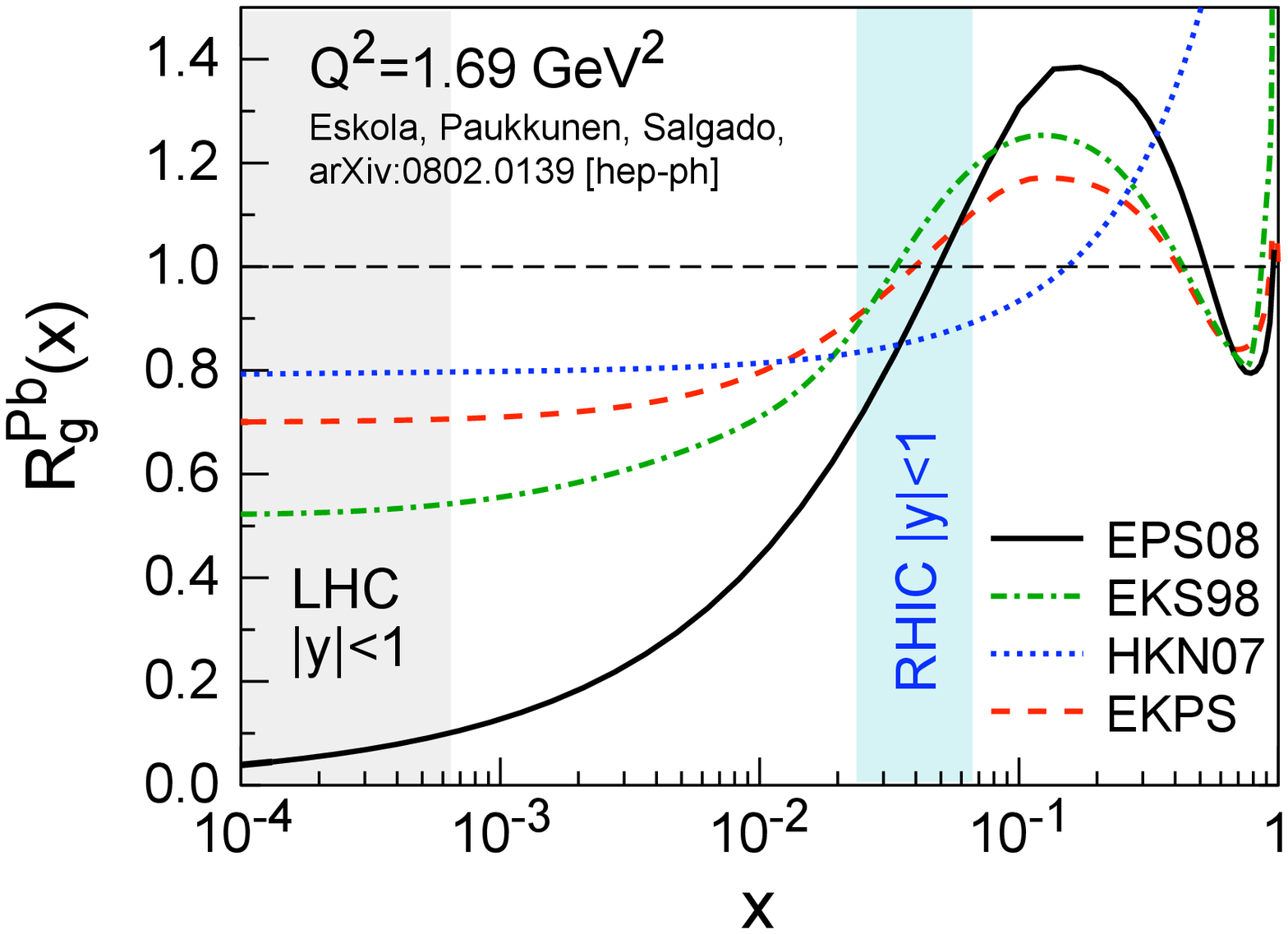}   
     \vspace{-5mm}
     \end{center}
     \caption{\textbf{Left:} Fractional contribution from $gg$, $qg$, and
         $qq$ scattering processes to $\pi^0$ production at
         mid-rapidity for RHIC (black) and LHC (blue)
         \cite{vogelsang}.  \textbf{Right:} Comparison of the gluon
         modification factors for gluons (Pb over p) from LO global
         DGLAP analyses \cite{Eskola:2008ca}.}
    \label{fig:werner_and_carlos}
\end{figure}

\textbf{Particle Production and Nuclear Effects}. 
The use of hard
probes to study the properties of hot matter in heavy ion collisions
is moving into the precision stage with high luminosities at RHIC and
high energies at the LHC.  
The strong suppression of high \pT\ hadrons
observed at RHIC is interpreted in terms of partonic energy
loss via induced gluon radiation in the high-density matter.  The
initial parton distributions (which determine the incoming flux) play
a crucial role in quantitatively extracting the amount of energy lost.
Figure \ref{fig:werner_and_carlos} (left) illustrates the role of
gluons in $\pi^0$ production in \pp\ collisions at
RHIC and LHC over a wide range of \pT\ \cite{vogelsang}. The
underlying NLO calculations \cite{nlo-wv} are based on established parton
distribution (CTEQ6M) and fragmentation (DSS) functions.  These
distributions are strongly modified in nuclei, with shadowing and
saturation at low $x$ and the EMC effect at moderate $x$.  To
calibrate the nuclear parton distributions, $G_A$ must be well
constrained for $x \geq 10^{-2}$ at RHIC, and $x \geq 10^{-4}$ at the
LHC for $Q^2 \sim $1--10\,GeV$^2$.  Figure \ref{fig:werner_and_carlos}
right shows that the current uncertainties, especially at the LHC, are
large, leading to differences in the final transverse energy flow by
factors of 2--4 and an order of magnitude uncertainty in semi--hard
cross-sections \cite{Eskola:2008ca}.  At the same time, RHIC data on
$\pi^0$ production in deuteron-gold collisions at high $p_T$
\cite{Adams:2006uz} and on photon production in \AA\ collisions at
high $p_T$ \cite{Miki_2008} suggest non-trivial modifications of
parton distributions at $x > 0.1$.  Precision measurements in the
kinematic regime relevant to the RHIC and LHC measurements are
essential for using hard probes to diagnose the active degrees of
freedom of the sQGP.

\textbf{Energy Loss and Hadronization in Hot Matter}.  While the RHIC
data is broadly explained by the attenuation of quarks and gluons in a
hot medium, quantitative studies require that the role of the cold
nuclear medium on partons and hadrons be well understood.  HERMES DIS
data \cite{Airapetian:2007vu} confirm that the energy loss and
\pT-broadening of formed hadrons produced in \eA\ collisions are
small, but the luminosity at HERMES is too low to study the
attenuation of charm or bottom quarks. The surprisingly large energy
loss of heavy quarks at RHIC poses major challenges to
theory~\cite{Djordjevic:2008iz}; conjectures about the role of
collisional energy loss and pre--hadron absorption in the attenuation
of heavy quarks can be tested, for the first time, in cold matter with
EIC.  Furthermore, the wide range of photon energies at an EIC 
($10\,\gev < \nu < 1600\,\gev$) compared to HERMES (2--25\,GeV)
offers more channels to study hadronization inside and outside of the
nucleus and to test the factorization (\eA/\pA/\ep/\pp) of the
fragmentation of partons into hadrons, especially with processes
involving heavy quarks. The wide range of photon energies
available at the EIC is
especially relevant as a cold matter benchmark for final states in
\AA\ collisions at the LHC, where the typical energies of jets will be
well above the maximal values available at HERMES.

\begin{figure}[tb]
     \begin{center}
     \includegraphics[width=0.47\textwidth]{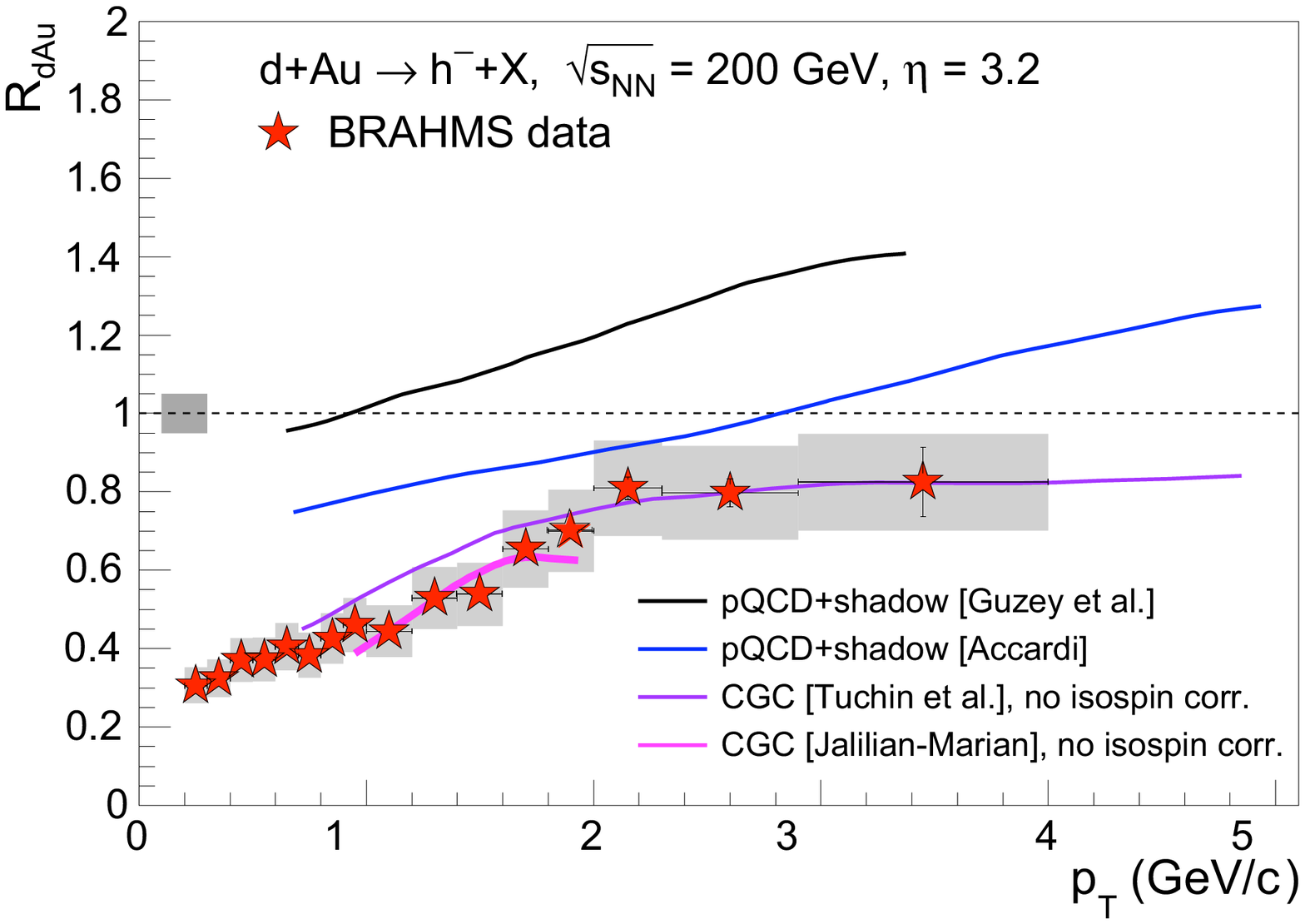}   
     \hspace{0.07\textwidth}
     \includegraphics[width=0.42\textwidth]{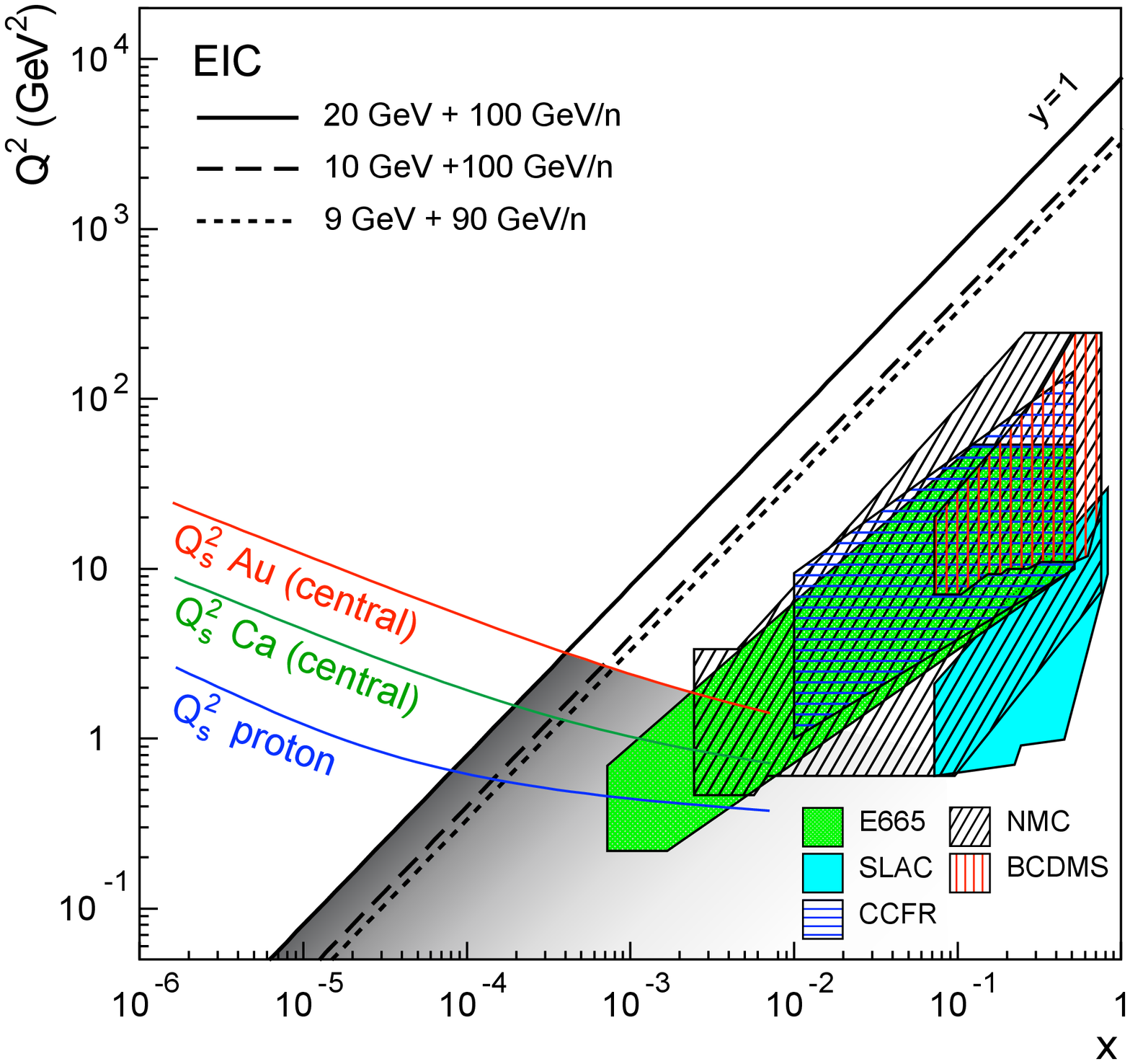}   
     \vspace{-5mm}
     \end{center}
     \caption{\textbf{Left:} $R_{\mathrm{dAu}}$ for negative charged 
     hadrons at forward
         pseudorapidity \cite{Arsene:2004ux}. \textbf{Right:}
         Kinematic acceptance in the ($Q^2,x$) plane for the EIC.
         Shown are lines for various complementary concepts to realize
         EIC.  Lines showing the quark saturation scale $Q_s^2$ for
         protons, Ca, and Au nuclei are superposed on the kinematic
         acceptance.  The shaded region indicates the range where
         saturation effects are to be expected.  The kinematic
         coverage of past $e$+A, $\mu$+A, and $\nu$+A experiments is
         indicated.}
    \label{fig:brahms}
\end{figure}

\textbf{Saturation Effects in the Forward Region at RHIC and at
    Midrapidity at LHC}.  Yields of moderate \pT\ particles
(2--4\,GeV) in the forward region ($\eta \approx 3.2$) of \dAu\ 
collisions at RHIC show a systematic suppression as the deuteron
passes through thicker regions of the Au nucleus, as shown in
Fig.~\ref{fig:brahms} (left).  These particles correspond to partons
of very low $x \approx {\cal{O}}(10^{-3} - 10^{-4})$, suggestive of
the relevance of saturation effects, especially with the large values
of ($\Qss \approx $2.5--5\,GeV$^2$) in this region.  These values are
comparable to those at mid-rapidity at the LHC.  The theory curves
shown correspond to different model assumptions. In the forward region
at the LHC ($y=3$), one expects saturation momenta of order $Q_s^2 =
10$\,GeV$^2$.  The establishment of saturation effects via
measurements of $G_A$ at the EIC will be vital to interpret
measurements in the forward region at RHIC and at all rapidities at
the LHC.

\section{EIC Accelerator Concepts}

The requirements for an \eA\ collider are driven by the need to access
the relevant region in $x$ and $Q^2$ that will allow us to explore
saturation phenomena in great detail. This region is defined by our
current understanding of $Q_s(x,Q^2)$ depicted in
Fig.~\ref{fig:brahms} (right). A machine that would reach sufficiently
large $Q \approx Q_s$ values in \ep\ collision would require energies
that are beyond current budget constraints. However, as pointed
already, at fixed $x$, $Q_s$ scales approximately with A$^{1/3}$.
Ions with large masses thus allows us to reach into the saturation
regime at sufficiently large Q values. To fully explore the physics
capabilities in \eA, double differential measurements at varying
\sqrts\ are mandatory. This can be only achieved if the provided beams
have large luminosities.

From these considirations the following requirements for an \eA\ 
collider evolve:

\begin{itemize}
\item Collisions of at least $\sqrt{s} >
    60$\,GeV to go well beyond the range explored in past fixed target
    experiments (see Fig.~\ref{fig:brahms}).
\item Luminosities of $L > 10^{30}$ \lumi, are required for
    precise and definitive measurements of the gluon distributions of
    interest.
\item Provision of ion beams at different
    energies which are mandatory for the
    study of many relevant distributions such as $F_L$.
\item Provision of a wide range of ions. For saturation
    physics studies, beams of very high mass numbers are vital.
\end{itemize}

\begin{figure}[bt]
     \begin{center}
    \includegraphics[width=0.5\textwidth]{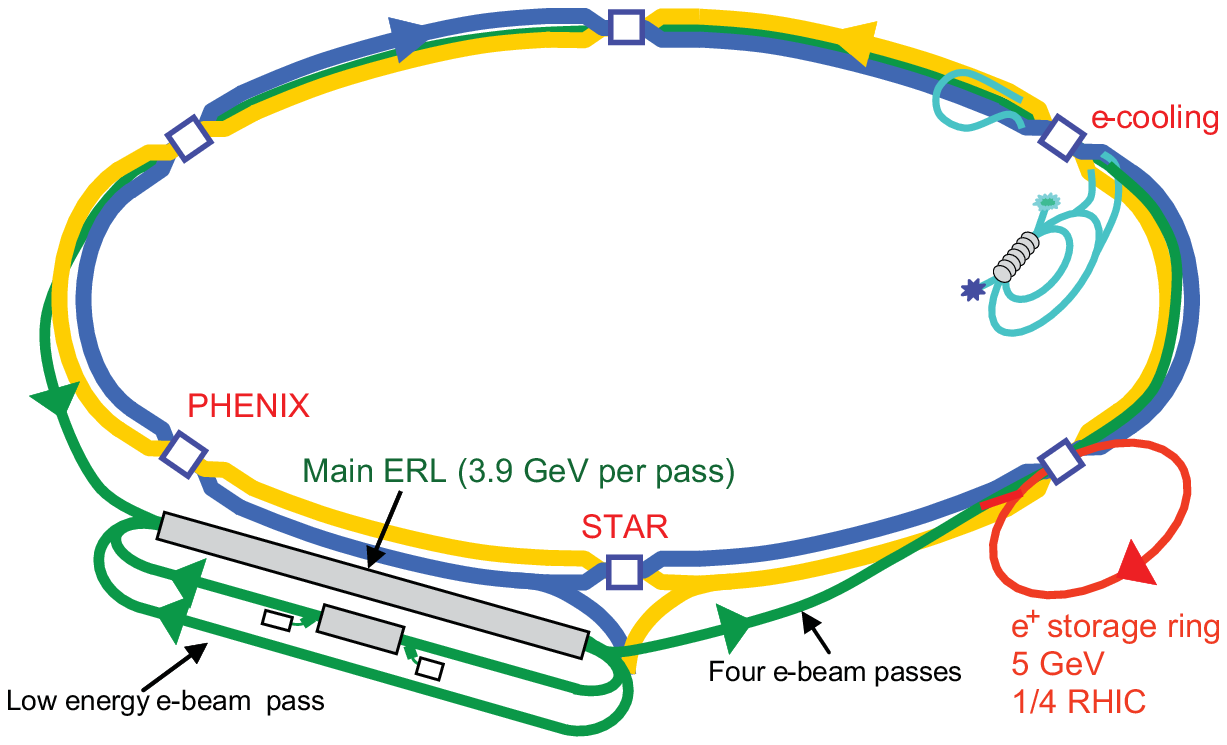}
    \hspace{0.05\textwidth}
    \includegraphics[width=0.38\textwidth]{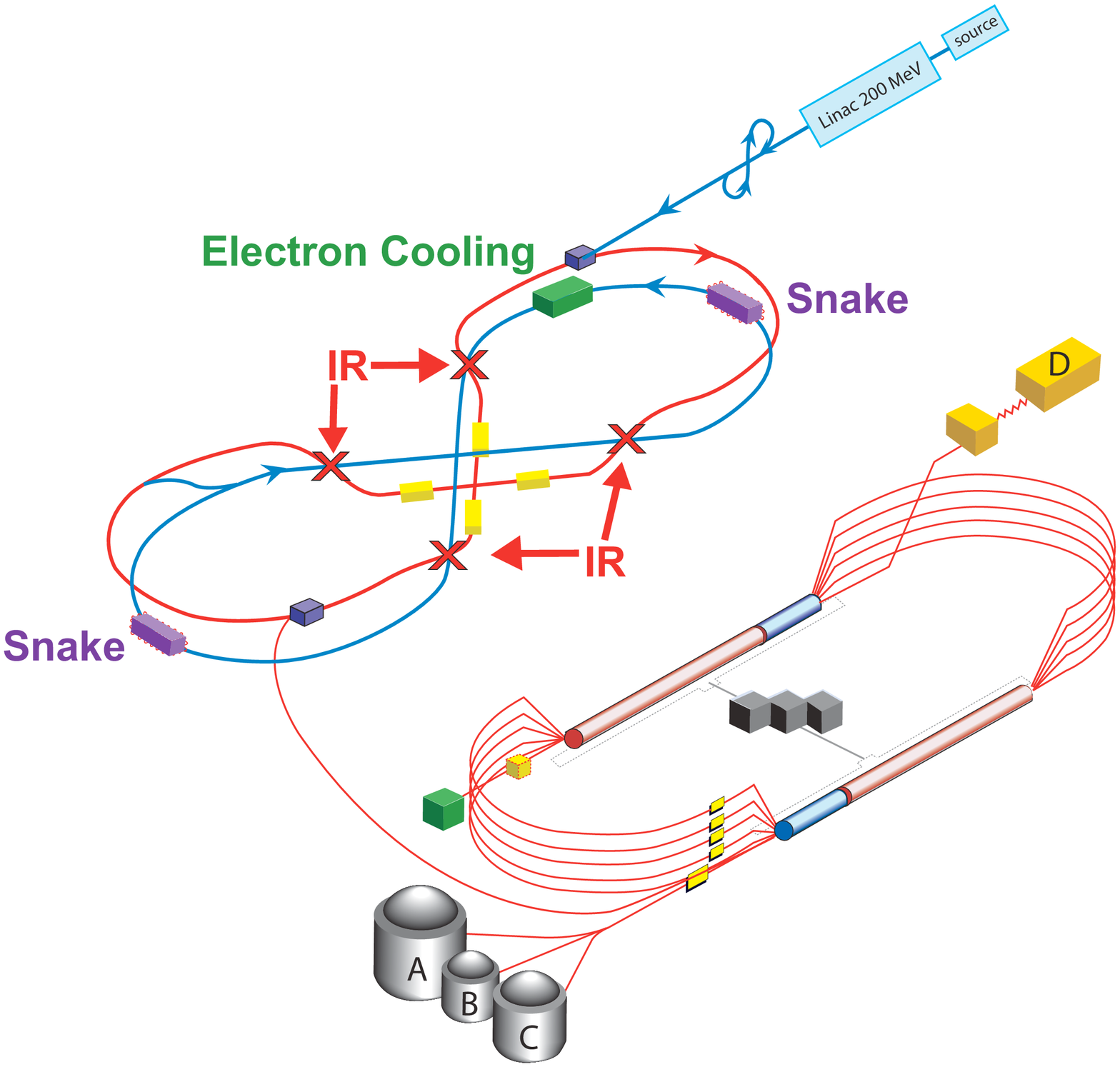}
    \vspace{-3mm}
     \end{center}
     \caption{Design layout of the eRHIC collider at BNL
         based on the Energy Recovery Linac (left)
         and the ELIC schematic layout (right) at the JLAB.}
    \label{fig:eic_concepts}
\end{figure}

There are currently two complementary concepts to realize an EIC:
eRHIC, which calls for the construction of a new electron beam to
collide with the existing RHIC ion beam; and ELIC, which calls for the
construction of a new ion beam to collide with the upgraded CEBAF
accelerator.  Both rely on new accelerator and detector technology,
and on an allocation of suitable R\&D resources for their expeditious
development.

For eRHIC, the most promising design option is based on the addition
of a superconducting energy recovery linac to the existing RHIC ion
machine \cite{vigdor2008}. The linac will provide the electron beam
for the collisions with ions or protons, circulating in one of the
RHIC rings.  The general layout of the machine is shown in
Fig.~\ref{fig:eic_concepts} (left). Design luminosity is $\sim 3\cdot
10^{33}$ \lumi\ for 20 GeV electrons on 100 GeV/n Au beams.

ELIC is envisioned as a future upgrade of CEBAF, beyond the planned 12
GeV upgrade for fixed target experiments. The CEBAF accelerator will
be used as a full energy injector into an electron storage ring.  An
new ion complex will be used to generate, accelerate, and store
polarized ions and unpolarized medium to heavy ions. Figure
\ref{fig:eic_concepts} (right) displays the conceptual layout of ELIC
at CEBAF. The design efforts aims for a peak luminosity of $10^{35}$
\lumi\ for 7 GeV electron on 75 GeV/n ions.

\section{Key Measurements in \textit{e}+A}

The \eA\ program at an EIC can be defined by a well defined list of 
measurements that address a set of specific key questions. They all
focus on the study of the properties of the gluon dominated region in nuclei.

\begin{figure}[bt]
    \begin{center}
    \includegraphics[width=0.7\textwidth]{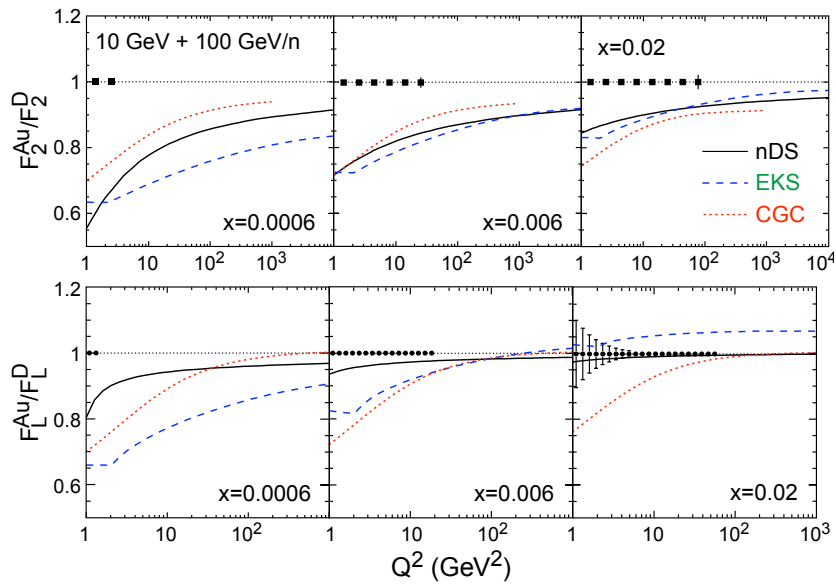}
    \end{center}
    \vspace{-5mm}
    \caption{The ratio of the structure function $F_2$ (upper row)
         and $F_L$ (lower row) in Au relative to the corresponding
         structure functions in the deuteron as a function of $Q^2$ for
         several bins in $x$. The filled circles and error bars
         correspond respectively to the estimated kinematic reach and
         the statistical uncertainties for a luminosity of 4/A fb$^{-1}$
         with the EIC for Au and d, respectively. For $F_L$, 3 runs at
         different $\sqrt{s}$ were used. The acronyms nDS and EKS stand for
         different sets of PDFs. The CGC predictions are only applicable
         at small $x$. }
\label{fig:fcombo}
\end{figure}

\textbf{What is the momentum distribution of gluons (and sea quarks) in
    nuclei?}  This is the measurement at the EIC that will lay the
foundation for further studies.  There are various techniques to
extract $G^A(x, Q^2)$: \textit{(i)} through scaling violations of
$F_2^A$ ($\partial F_2^A/\partial \ln(Q^2) \neq 0$) with $Q^2$,
\textit{(ii)} through the structure function $F_L^A$ which is directly
proportional to the gluon distribution in the framework of pQCD, and
\textit{(iii)} through the measurement of inelastic and \textit{(iv)}
diffractive vector meson production. Figure \ref{fig:fcombo} shows
projections of EIC measurements of $F_2^A$ and $F_L^A$ that encode the
parton density information, plotted as ratios of the values measured
for a gold nucleus to that for deuterium. The statistical precision
attainable in just $\sim 10$ weeks of running at readily achievable
collider luminosities will allow EIC data to distinguish clearly
between the saturated gluon densities associated with the CGC and
those anticipated in linear QCD approaches.
    
\textbf{What is the space-time distribution of gluons (and sea quarks)
    in nuclei?}  The nature of the spatial distribution of glue
provides a unique handle on the physics of high parton densities and
has important ramifications for a wide range of final states in
hadronic and nuclear collisions. Is the ``gluon density profile" in
the nucleus in the transverse plane one of small clumps of glue or is
it more uniform?  Techniques to extract information about the spatial
distribution of glue based on the measurement of vector meson
production (\textit{e.g.} $\rho$ and $J/\psi$) were developed for \ep\ 
at HERA (see for example \cite{Munier:2001nr,Rogers:2003vi}) and are
directly applicable to \eA.
        
\textbf{What is the role of color neutral (Pomeron) excitations in
    scattering off nuclei?}  Diffractive interactions result when the
electron probe in DIS interacts with a color neutral vacuum
excitation, the Pomeron. At HERA, an unexpected discovery was that
~15\% of the \ep\ cross-section is from diffractive final states.
This is a striking result implying that a proton at rest remains
intact one seventh of the time when struck by a 25\,TeV electron.  The
effect is even more dramatic in nuclei. Several models of strong gluon
fields in nuclei suggest that large nuclei are intact $\sim$ 25-30\%
of the time \cite{Kowalski:2007rw}.  Measurements of coherent
diffractive scattering on nuclei are easier in the collider
environment of EIC relative to fixed target experiments.  Studies at
the EIC will allow to directly probe the nature of the Pomeron and
further will provide definitive tests of strong gluon field dynamics
in QCD.
       
\textbf{How do fast probes interact with an extended gluonic medium?}
In nuclear DIS one observes a suppression of hadron production
\cite{Airapetian:2007vu, Accardi:2007in} analogous to, but weaker than
in, heavy-ion collision at RHIC
\cite{Adcox:2004mh,Adams:2005dq,Back:2004je,Arsene:2004fa}.  It
provides the cleanest environment to address nuclear modifications of
hadron production. One can experimentally control many kinematic
variables; the nucleons act as femtometer-scale detectors allowing one
to experimentally study the propagation of a parton in this ``cold
nuclear matter'' and its space-time evolution into the observed
hadron.

Some of these measurements described in this section have analogs in \ep\ 
collisions but have never been performed in nuclei; for these, \eA\ 
collisions will allow us to understand universal features of the
physics of the nucleon and the physics of nuclei. Other measurements
have no analog in \ep\ collisions and nuclei provide a completely
unique environment to explore these.

\section{Connection to \textit{p}+A Physics}

Both \pA\ and \eA\ collisions can provide excellent information on the
properties of gluons in the nuclear wave functions. Only DIS, however,
allows the direct determination of the momentum fraction $x$ carried
by the struck parton before the scattering and the momentum $Q$
transferred to the parton in the scattering process and thus a precise
mapping of $G(x, Q^2)$.

Deeply inelastic \eA\ collisions are dominated by one photon exchange;
they have a better chance to preserve the properties of partons in the
nuclear wave functions because there is no direct color interaction
between $e$ and A.  The photon could interact with one parton to probe
parton distributions, as well as multiple partons coherently to probe
multi-parton quantum correlations \cite{Qiu:2003vd}.

Many observables in \pA\ collisions require gluons to contribute at
the leading order in partonic scattering. Thus \pA\ collisions provide
more direct information on the response of a nuclear medium to a gluon
probe. However, soft color interactions between $p$ and A before the
hard collision takes place have the potential to alter the nuclear wave
function and destroy the universality of parton properties
\cite{Collins:gx}.  Such soft interactions contribute to physical
observables as a correction at order $1/Q^4$ or higher
\cite{Doria:1980ak,Di'Lieto:1980dt,Qiu:1990xy}.  These power
corrections cannot be expressed in terms of universal parton
properties in the nuclear wave functions, thereby breaking QCD
factorization. The breakdown of factorization has already been
observed in comparisons of diffractive final states in \ep\ collisions
at HERA and \pp\ collisions at the Tevatron \cite{Schilling:2002tz}.

Due to the very large reach in $x$ and $M^2$, \pA\ collisions at the
LHC have significant discovery potential for the physics of strong
color fields in QCD. However, due to uncertainties relating to
convolutions over parton distributions in the proton probe, final
state fragmentation effects, and factorization breaking contributions,
the results can be cleanly interpreted only for $M^2\gg Q_s^2$ where
the strong field effects will be weaker.

\section{Summary}

Precision measurements with an EIC will open a new window to the
regime dominated by direct manifestations of the defining feature of
QCD: gluons and their self-interactions. These self-interactions lie
at the heart of nucleon and nuclear structure and are expected to be
essential to the understanding of high energy heavy-ion collisions.
To date, their properties and dynamics in matter remain largely
unexplored.  A high luminosity EIC with center-of-mass energy in the
range from 30 to 100 GeV with polarized nucleon beams and the full
mass range of nuclear beams can be realized either at RHIC or at JLAB.
It will provide access to those low-$x$ regions in the nucleon and
nuclei where their structure is governed by gluons. In addition,
polarized beams in the EIC will give unprecedented access to the
spatial and spin structure of gluons in the proton. While significant
progress has been made in developing concepts for an EIC, many open
questions remain. Realization of an EIC will require essential R\&D in
a number of areas including: cooling of high-energy hadron beams, high
intensity polarized electron sources, and high energy, high current
Energy Recovery Linacs.  Over the next five years significant progress
must be made in these areas and the community has to converge on one,
optimized design for the accelerator. The EIC would provide unique
capabilities for the study of QCD well beyond those available at
existing facilities worldwide and complementary to those planned for
the next generation of accelerators in Europe and Asia.

\section*{Acknowledgment}

I am grateful to my BNL colleagues Raju Venugopalan, Werner Vogelsang,
and Dave Morrison for discussions and their
help in preparing this manuscript. This
work was supported by the U.S. Department of Energy under Grant No.
DE-AC02-98CH10886.

\end{document}